\documentclass[floats,floatfix,showpacs,amssymb,prd,twocolumn,superscriptaddress,nofootinbib,nolongbibliography,reprint]{revtex4-2}

\usepackage{amssymb,amsmath,mathtools,needspace,enumitem,etoolbox,graphicx,physics,microtype,afterpage,bigints,gensymb,tabularx,bigints,relsize,soul}

\usepackage[dvipsnames, usenames]{xcolor}
\definecolor{linkcolor}{rgb}{0.0,0.3,0.5}
\usepackage[unicode, colorlinks=true, linkcolor=linkcolor, citecolor=linkcolor, filecolor=linkcolor,urlcolor=linkcolor, pdfusetitle]{hyperref}
\usepackage[all]{hypcap}
\usepackage[T1]{fontenc}
\usepackage[utf8]{inputenc}
\usepackage{orcidlink}
\usepackage[thinc]{esdiff}

\usepackage[normalem]{ulem}

\newcommand{\ssim}{\mathchar"5218\relax\,}

\makeatletter
\newcommand*{\balancecolsandclearpage}{\close@column@grid \cleardoublepage \twocolumngrid}
\makeatother

\newcommand{\milan}{\affiliation{Dipartimento di Fisica ``G. Occhialini'', Universit\'a degli Studi di Milano-Bicocca, Piazza della Scienza 3, 20126 Milano, Italy}}
\newcommand{\infn}{\affiliation{INFN, Sezione di Milano-Bicocca, Piazza della Scienza 3, 20126 Milano, Italy}}
\newcommand{\AEI}{\affiliation{Max Planck Institute for Gravitational Physics (Albert Einstein Institute) Am M\"uhlenberg 1, 14476 Potsdam, Germany}}
\newcommand{\MIT}{
\affiliation{LIGO, Massachusetts Institute of Technology, 77 Massachusetts Avenue, Cambridge, MA 02139, USA}
\affiliation{Kavli Institute for Astrophysics and Space Research \& Department of Physics, \\
Massachusetts Institute of Technology, 77 Massachusetts Avenue, Cambridge, MA 02139, USA}}
\newcommand{\nott}{\affiliation{School of Mathematical Sciences, University of Nottingham,
University Park, Nottingham NG7 2RD, United Kingdom}}

\begin{document}

\title{Reconstructing parametric gravitational-wave population fits \\ from nonparametric results without refitting the data}

\author{Cecilia Maria Fabbri$\,$\orcidlink{0000-0001-9453-4836}}
\email{cecilia.fabbri@nottingham.ac.uk}
\milan \nott

\author{Davide Gerosa$\,$\orcidlink{0000-0002-0933-3579}}
\milan \infn

\author{Alessandro Santini$\,$\orcidlink{0000-0001-6936-8581}}
\AEI 

\author{\\Matthew Mould$\,$\orcidlink{0000-0001-5460-2910}}
\MIT

\author{Alexandre Toubiana$\,$\orcidlink{0000-0002-2685-1538}}
\milan \infn

\author{Jonathan Gair$\,$\orcidlink{0000-0002-1671-3668}}
\AEI 

\pacs{}

\date{\today}

\begin{abstract}

Combining multiple  events into population analyses is a cornerstone of gravitational-wave astronomy. A critical component of such studies is the assumed population model, which can range from %
astrophysically motivated functional forms to nonparametric treatments that are flexible but difficult to interpret. In practice, the current approach is to fit the data multiple times with different population models to identify robust features.
We propose an alternative strategy: assuming the data have already been fit %
with a 
flexible model, we present a practical recipe to reconstruct the %
population distribution of a different %
model. As our procedure postprocesses existing results, it avoids the need to access the underlying gravitational-wave data again and handle selection effects. Additionally, our reconstruction metric provides a goodness-of-fit measure to compare multiple models.
We apply this method to the mass distribution of black-hole binaries detected by LIGO/Virgo/KAGRA.
Our work paves the way for streamlined gravitational-wave population analyses by first fitting the data %
with advanced nonparametric methods and careful handling of selection effects,  
while the astrophysical interpretation is then made accessible using our reconstruction procedure on targeted models.
The key principle is that of conceptually separating data description from data interpretation.

\end{abstract}

\maketitle

\section{Introduction}

With observational gravitational-wave (GW) astronomy now entering its teenage years, the dataset of merging black holes (BHs) and neutron stars counts approximately 100 events \citep{2019PhRvX...9c1040A,2021PhRvX..11b1053A,2023PhRvX..13d1039A,2024PhRvD.109b2001A}. By the next 10-year anniversary, the number of detections is expected to reach thousands, if not millions \citep{2018LRR....21....3A,2010CQGra..27s4002P,2019BAAS...51g..35R,2024arXiv240207571C}.
Such a population of merging objects is being dissected to reveal its many features \citep{2024arXiv241019145C}, and as the number of detected sources grows, more and more details %
will become  accessible.

The analysis of GW data typically proceeds in three steps \cite{2024arXiv240902037C}. First, stretches of data $d_i$ containing astrophysical signals are identified (``search''). These are then analyzed individually using Bayesian methods (``parameter estimation''), yielding posterior distributions $p(\theta | d_i)$ of the source parameters $\theta$, such as masses, spins, and redshifts. Finally, information from multiple events is combined into a ``population analysis,'' aimed at characterizing hyperparameters $\lambda$ that describe the entire set of detections \cite{2019MNRAS.486.1086M,2022hgwa.bookE..45V}.
This final step is conducted within the framework of hierarchical Bayesian statistics, which assumes a generating process —commonly an inhomogeneous Poisson process— and applies an astrophysical prior $p_{\rm pop}(\theta | \lambda)$ on the underlying distribution of source properties. Crucially, selection effects must be accounted for to infer the intrinsic distribution of sources from the limited subset that was detected.

GW-population results inevitably depend on the adopted model, to the extent that developing suitable expressions for $p_{\rm pop}(\theta | \lambda)$ has become a major focus in the field. Broadly speaking, three main strategies exist, though the boundaries between them are often blurred:

\begin{enumerate}

\item One approach is to assume simple functional forms (e.g., power laws, Gaussians) inspired by the underlying astrophysics. For instance, the stellar initial-mass function is well described by a power law \cite{2002Sci...295...82K}, so it is not unreasonable to look for a power-law structure in the BH mass spectrum as well. These parametric models are still considered state of the art \cite{2023PhRvX..13a1048A}. From here, one can either move closer to the data, or closer to the astrophysics.   

\item There have been several promising attempts at modeling GW populations with nonparametric, data-driven methods \cite{2024PhRvX..14b1005C,2023ApJ...946...16E,2017MNRAS.465.3254M,2023ApJ...957...37R,2022MNRAS.509.5454R,2023MNRAS.524.5844T,2025PhRvD.111f3043H,2024arXiv241023541N,2022PhRvD.105l3014S,2024ApJ...960...65S}. In this framework, the goal is to provide an accurate description of the population without necessarily linking findings to the astrophysical origins of the sources. The hyperparameters $\lambda$ in these models are often abstract (e.g., spline nodes, bin counts, or concentration parameters) and lack direct physical interpretability.

\item Alternatively, $p_{\rm pop}(\theta|\lambda)$ can be modeled using astrophysical simulations \cite{2008ApJS..174..223B,2010ApJ...719..915C,2022ApJS..258...34R,2020ApJ...898...71B,2018MNRAS.480.2011G,2024PhRvD.110d3023K}. In this case, the hyperparameters $\lambda$ correspond to the input parameters of the adopted population-synthesis code, such as supernova kick velocities, common-envelope efficiency, or stellar wind prescriptions. While this allows for direct constraints on the targeted physical processes, it results in an interpretation of the data that is strongly model dependent. 
Several works have proposed methods to solve the computational challenge of running many simulations using surrogate models, allowing for direct inference of astrophysical parameters from GW data \cite{2018PhRvD..98h3017T,2019PhRvD.100h3015W,2021PhRvD.103h3021W,2022PhRvD.106j3013M}.

\end{enumerate}

Identifying the ideal balance between flexibility and interpretability remains an open problem. In practice, the current approach is to fit the data multiple times using different population models $p_{\rm pop}(\theta | \lambda)$, ideally covering all strategies (i)–(iii) outlined above, and then compare the results \cite{2023PhRvX..13a1048A}. While %
insightful, %
this approach can be computationally expensive and might become impractical, particularly when handling selection effects.
Selection effects are typically modeled using importance sampling from a fixed set of software injections into search pipelines~\cite{2018CQGra..35n5009T,2019RNAAS...3...66F}. However, the injected distribution might not be equally suited for all choices of $p_{\rm pop}(\theta | \lambda)$. %

Here we propose a different approach, namely that of separating the description of the data and their interpretation. We present a post-processing procedure to transform results obtained under one population model into those associated with alternative models. 
In our framework, data would {first} be analyzed %
 {using a state-of-the-art} nonparametric method that carefully accounts for selection effects. This initial analysis would serve as a %
 starting point, %
 enabling the strategy outlined here to reconstruct the results for various alternative population models. %
 These output models could prioritize interpretability, bringing them closer to the questions of astrophysical interest.

Crucially, our
conversion procedure operates in the %
intrinsic (not observed) %
space of source parameters, ensuring that selection effects do not need to be accounted for again. The potential impact of this approach is substantial: data analysts who work closely with the detectors and possess the requisite expertise in their complexities are best positioned to deliver an accurate description of the data. Astrophysicists can then use these results to explore multiple models, testing the inclusion or exclusion of specific physical processes and assessing their effects.

\section{Methods}
\label{methods}

\subsection{Loss function}\label{loss}

Let us consider two population assumptions:\footnote{While we use the same symbol $p_{\rm pop}$ for both models, it is intended that their functional forms may differ. For simplicity, this distinction is encoded solely in the symbols used for the hyperparameters, $\gamma$ and $\lambda$. %
} an input model $p_{\rm pop}(\theta|\gamma)$ and an output model $p_{\rm pop}(\theta|\lambda)$. In our proposed framework, the hyperparameters $\gamma$ describe a flexible nonparametric model, while $\lambda$ corresponds to an interpretable astrophysical model.
Suppose the input model has already been fitted to the data $d$, and we have access to posterior samples $\gamma_i \sim p(\gamma|d)$, such as those routinely released by the LIGO--Virgo--KAGRA (LVK) Collaboration \cite{2023ApJS..267...29A}. Our goal is to post-process these samples to approximate the population posterior $p(\lambda|d)$ under the output model.

The strategy we present operates on a sample-by-sample basis. For each $\gamma_i$, we aim to identify the corresponding $\lambda_i$ such that the probabilities $p_{\rm pop}(\theta|\gamma_i)$ and $p_{\rm pop}(\theta|\lambda_i)$ are as similar as possible. This ensures that the transformation preserves the features of the population described by the input model while reinterpreting them in terms of the output model.

Implementing this idea requires a notion of distance between probability distributions, which is not unique~\cite{chung1989measures}. A widely used choice is the Kullback-Leibler divergence, or entropy, given by
\begin{equation} \textrm{KL}\big[p(x),q(x)\big] = \int p(x)\log{\frac{p(x)}{q(x)}} \,\textrm{d}x \,, \end{equation} which has strong foundations in probability theory and is commonly interpreted as the information gain when moving from $q(x)$ to $p(x)$~\cite{2020arXiv200302030L}.
In the following, we adopt a symmetrized version of the Kullback-Leibler divergence known as 
the Jeffreys divergence~\cite{2019Entrp..21..485N}: 
\begin{align}
	\textrm{J}\big[p(x),q(x)\big] &= \textrm{KL}\big[p(x),q(x)\big] + \textrm{KL}\big[q(x),p(x)\big] 
	\notag \\
	&=\int [p(x)-q(x)]\log{\frac{p(x)}{q(x)}} \textrm{d}x\,.
\end{align}

For each input sample $\gamma_i$, we frame the problem as a multi-dimensional optimization in the $\lambda$ space, minimizing the following loss function:
\begin{align}
{\rm loss} (\lambda) = \textrm{J} \big[p_{\rm pop}(\theta|\lambda), p_{\rm pop}(\theta|\gamma_{i})\big]\,.
\label{lossf}
\end{align}
This loss function has the desirable property of reducing the optimization to separate subspaces in the case of factorized population models.
For example, many current analyses in GW astronomy partition both the parameters $\theta$ and the hyperparameters $\lambda$ into three distinct groups: those related to the BH masses $m$, those related to the BH spins $\chi$, and those related to the source redshift $z$,
\begin{align}
\theta = \{\theta_{m}, \theta_{\chi}, \theta_{z}\}\,, \; \; \; 
\lambda = \{\lambda_{m}, \lambda_{\chi}, \lambda_{z}\}.
\end{align}
The population model is then written as 
\begin{align}
p_{\rm pop}(\theta|\lambda) = p_{\textrm{pop},m} (\theta_{m}|\lambda_{m})\, p_{\textrm{pop},\chi}(\theta_{\chi}|\lambda_{\chi}) 
\,p_{\textrm{pop},z}(\theta_{z}|\lambda_{z})\,.
\label{eq: ppop_factorizes}
\end{align}
In the specific case where this factorization holds for both the input and output models, the loss function from Eq.~(\ref{lossf}) can be written as
\begin{equation}
{\rm loss} (\lambda)  = {\rm loss} (\lambda_m) +{\rm loss} (\lambda_\chi) +{\rm loss} (\lambda_z) 
\label{eq: Jeff_separates}
\, ,
\end{equation}
where 
\begin{equation}
{\rm loss} (\lambda_X) = \textrm{J} \big[p_{\rm pop,X}(\theta_X|\lambda_{X}), p_{\rm pop,X}(\theta_X|\gamma_{X,i})\big] %
\label{losseq}
\end{equation}
for $X=\{m,\chi,z\}$. %
The higher-dimensional optimization problem in the $\lambda$-space can thus be tackled separately in each of the lower-dimensional subspaces. Moreover, if the input and output prescriptions share some submodels (for example, if the spin and redshift assumptions are the same and the goal is to only reconstruct the mass distribution), the optimal solution in these common subspaces is trivially given by $\lambda_{X} = \gamma_{X}$. Though the loss function is not a unique choice, this highly desirable property is the key reason for choosing the Jeffreys divergence.

\subsection{Interpretation}
\label{interpretation}

The procedure we just highlighted is not expected to {exactly} recover the posterior that would be found from directly fitting the output model to the data.
 The reason is that all samples of the input model are treated equally in the posterior for the output model, irrespective of their minimized loss. 
In the following, we indicate the direct fit with $p(\lambda| d)$ and our approximation with $p(\lambda | d, {\rm Rec}_{\gamma})$, where $ {\rm Rec}_{\gamma}$ schematically denotes a reconstruction from a different model with some $\gamma$ hyperparameters. 

The input samples are distributed proportionally to the posterior $p(\gamma| d)$ under the flexible model, which in the data-relevant region of the parameter space reduces to the likelihood. Samples with small (large) minimized losses correspond to populations from the input model that are (not) well described by the output model; the likelihoods for the two models would therefore be expected to be similar (different). In the output-model posterior $p(\lambda| d)$, the samples would instead be distributed according to the likelihood for the output model and so we expect to see some differences. 

It might be expected that the approximate posterior $p(\lambda | d, {\rm Rec}_{\gamma})$ would tend to be broader than the directly fitted posterior $p(\lambda| d)$, since it will include additional samples with high minimized loss. This is not a quantitative statement, however, and the results presented below do not strongly support this expectation, at least for the cases explored here. Regardless, significant differences between $p(\lambda | d)$ and $p(\lambda | d, {\rm Rec}_{\gamma})$ will arise only if there are samples in the input posterior that look very unlike populations in the output model. As the input model is flexible, this would mean the output model is not a good description of the underlying population and therefore we should not be interested in obtaining a posterior for it. As discussed later in Sec.~\ref{goodness}, our method includes a diagnostic of the quality of the output model that can be used to identify such cases.

\subsection{Numerical implementation}
\label{Numerical implementation}

We minimize the loss function of Eq.~(\ref{lossf}) using simulated annealing \cite{{1997PhLA..233..216X}, {1988JSP....52..479T},{1996PhyA..233..395T},{2000PhRvE..62.4473X}}, which is an algorithm specifically designed to approximate global minima in high-dimensional spaces. %
We use the \textsc{scipy} implementation \cite{2020NatMe..17..261V}.

Although global optimization algorithms %
are less likely to get stuck in local minima compared to optimizers based on, e.g., gradient descent, one can never mathematically prove that the true global minimum has been identified. To mitigate this, we run the algorithm multiple times using different initial guesses and select the best solution among all attempts. (i) We first initialize the search in a random location extracted uniformly in the targeted hyperparameter range. %
From this first optimization, we have one value of $\lambda_{i}$ for each of the $\gamma_{i}$'s. After excluding samples that have at least one element coincident with a prior bound, we compute the median of the $\lambda_{i}$'s in each dimension, obtaining a single location in the $\lambda$ parameter space.
The goal here is to identify a value of $\lambda$ that provides a reasonable solution and thus a somewhat informed initial guess. 
(ii) We then run two iterations starting from around this location; more specifically, guesses are drawn from uniform distributions centered on the precomputed median value with half-interval equal to $5\%$ and $1\%$ of the prior range. We retain the $\lambda$ value that yields the minimum loss after this second step. We have tested this strategy on both toy models as well as the science cases described in Sec.~\ref{results} and found it to be robust.

The time required for each minimization is approximately 1.5 hours on a single CPU. While the computational cost of our method is currently higher than that of a full hierarchical Bayesian run, we emphasize that this is a first-attempt implementation, whereas hierarchical  sampling for GW astronomy has been extensively optimized over the past decade. Possible optimization strategies include just-in-time compilation and exploring alternative algorithms beyond simulated annealing.
It is also worth noting that the accuracy of a full hierarchical run depends on the size of the GW catalog, possibly scaling even quadratically with the number of events~\cite{2023MNRAS.526.3495T}. On the other hand, the computational effort involved in our postprocessing procedure does not depend on the number of detections but only on the number of hyperposterior samples.

\subsection{Hyperparameter ranges}
\label{validity}

When minimizing ${\rm loss}(\lambda)$, one must specify the region of the $\lambda$ space where solutions are to be sought. This is conceptually similar to defining prior ranges in stochastic sampling runs, effectively zeroing out the posterior outside the prescribed range.
When minimizing Eq.~(\ref{lossf}), there is no guarantee that the optimal solution will fall in the bulk of the imposed range; it could instead lie at the extrema of one or more dimensions in the $\lambda$ space. 

This situation is often referred to as ``railing against the prior'' in stochastic sampling and is typically addressed by rerunning the analysis with a wider prior range.
Instead of rerunning, in the application presented in Sec.~\ref{results} we opted to exclude such cases  from the reported output distributions.  This is to ensure meaningful comparisons with the dataproduct of  Ref.~\cite{2023PhRvX..13a1048A}, as changing the prior ranges would effectively result in comparisons against a different model. %

\subsection{Merger rates}
\label{secrates}

The hyperparameters $\lambda$ and $\gamma$ encode information exclusively about the shape of the source parameter distribution. Generalizing our reconstruction procedure to include detection rates is a straightforward extension.

The quantity entering the population likelihood is the number of BH mergers per parameter-space interval, ${\dd \mathcal{N}}/{\dd \theta}$ \cite{2019MNRAS.486.1086M,2022hgwa.bookE..45V}. Without loss of generality, this can be expressed as a product separating the normalization and the shape of the distribution:
\begin{equation}
\frac{\dd \mathcal{N}}{\dd \theta} (N_\lambda, \lambda) = N_\lambda\, p_{\rm pop}(\theta, \lambda),
\end{equation}
where $N_\lambda$ represents the total number of mergers, and $p_{\rm pop}(\theta, \lambda)$ is the normalized shape of the distribution, satisfying $\int p_{\rm pop}(\theta, \lambda) \dd \theta = 1$. The hyperparameters $N_\lambda$ and $\lambda$ are both inferred from the data.

When reconstructing an output model $(N_\lambda, \lambda)$ from an input fit $(N_\gamma, \gamma)$, the shape and normalization can be treated separately. The reconstructed normalization is trivially equal to the input normalization. Specifically, for each input sample $(N_{\gamma,i}, \gamma_i)$, the reconstructed output sample $(N_{\lambda,i}, \lambda_i)$ satisfies $N_{\lambda,i} = N_{\gamma,i}$, while $\lambda_i$ is determined using the minimization procedure outlined above.

A slight complication is that, most often, the quantity $N_\lambda$ is not provided directly.
Instead, a common approach is to parametrize the distribution using the local merger rate density $R_0$, measured in ${\rm Gpc}^3 \, {\rm yr}^{-1}$ \cite{2018ApJ...863L..41F,2023PhRvX..13a1048A}. The conversion between $R_0$ and $N_\lambda$ is straightforward but depends on the shape hyperparameters $\lambda$ (see, e.g., Ref.~\cite{2025PhRvD.111d4048D} for explicit expressions).
Starting with posterior samples of $R_0$ from the input model, one should first convert these to $N_\gamma$ using the corresponding $\gamma$ values, then set $N_\lambda = N_\gamma$, and finally convert back to $R_0$ using the reconstructed $\lambda$ values.

\section{Results}
\label{results}

\subsection{Adopted models}

We apply our reconstruction strategy to two state-of-the-art population models used by the LVK Collaboration in their GWTC-3 BH-binary population analysis~\cite{2023PhRvX..13a1048A}:
\begin{enumerate}
\item
 \textsc{Power Law+Peak} (PP)~\cite{2018ApJ...856..173T} is an explicitely parametric approach. The primary BH masses $m_1$ are assumed to follow a mixture of a power law and a Gaussian component; the former is modeled after the initial mass function of stars, while the latter is meant to capture a potential build-ups due to e.g. pair instability supernovae. The hyperparameters %
include the spectral index of the power law, the moments of the Gaussian, the mass cutoffs, and the mixing fraction between the two components. 

\item \textsc{Power Law+Spline} (PS)~\cite{2022ApJ...924..101E} is a semi-parametric approach. The BH primary masses are distributed according to a perturbed power-law distribution, where the perturbation is modeled using a spline function with a fixed set of nodes. %
This is a more flexible model, designed to identify multiple overdensities and underdensities. The hyperparameters %
of the models are the spectral index of the power law as well as the heights of the spline nodes. These are indeed quite abstract and lack a direct physical interpretation.

\end{enumerate}
For both models, the secondary mass $m_2$ is assumed to follow a power-law distribution. Note this is not independent of ---but rather conditioned on--- the value of $m_1$, i.e., $p_{\rm pop}(m_1,m_2| \lambda)= p_{\rm pop}(m_1| \lambda) p_{\rm pop}(m_2| m_1,\lambda)$.

This paper presents  
a reconstruction of the PP hyperposterior using PS  samples. While here we reconstruct one model from the other, both models were also fitted directly to the data \cite{2023PhRvX..13a1048A}, providing an ideal testing ground for our new procedure. 
For testing purposes, in the following we also present some results assuming a simpler power-law-only model (PL), which is defined as a subset of both the PP and the PS models where either the fraction of sources in the Gaussian component or the spline perturbation amplitudes are set to zero. %

Reference~\cite{2023PhRvX..13a1048A} considered these mass models without changing the spin and redshift parametrizations. This corresponds to the case described in Sec.~\ref{loss} where the population models are partitioned into multiple subsets. We can thus optimize
Eq.~(\ref{eq: Jeff_separates})
independently of
spin and redshift. %

We use 
the population model implementations from \textsc{GWPopulation} \cite{
2019PhRvD.100d3030T}
and
the public LVK data release supporting Ref.~\cite{2023PhRvX..13a1048A}, selecting version 1 for PS \cite{ligo_scientific_collaboration_and_virgo_v1} and version 2 for PP \cite{ligo_scientific_collaboration_and_virgo_v2}.
We use the same notation as Ref.~\cite{2023PhRvX..13a1048A}, to which we refer for the precise expression of the adopted models. The hyperparameters of the PP
parametrization are denoted by the symbols $\alpha, \beta_q, m_{\rm max}, m_{\rm min}, \lambda_{\rm peak}, \mu_m, \sigma_m$, and $\delta_m$. For the PS model, the spline perturbation amplitudes are denoted by $f_{i}$ with $i=0,\dots, 19$. 
In all cases, we consider the priors of Ref.~\cite{2023PhRvX..13a1048A} and force our optimization algorithm to look for solutions exclusively in those ranges; cf. Sec.~\ref{validity}. 

\subsection{\textsc{Power Law+Peak} $\to$ \textsc{Power Law+Peak}}
\label{sec: PP2PP}

First, we test our procedure using the PP model as both input and output. This is a trivial scenario, where each single optimized sample $\lambda_i$ should coincide with the corresponding input sample $\gamma_i$. %
 Quoting the medians %
 of the corresponding distributions, we find that the differences between the input and output hyperparameters across the available samples are %
$\Delta \alpha \sim 10^{-7}$,
$\Delta \beta_q \sim 10^{-7}$,
$\Delta m_{\rm max}\sim 10^{-4} M_\odot $,
$\Delta m_{\rm min} \sim 10^{-7} M_\odot $,
$\Delta \lambda_{\rm peak} \sim {10^{-9}}$,
$\Delta \mu_m \sim 10^{-6} M_\odot$,
$\Delta \sigma_m \sim 10^{-7} M_\odot$,
and
$\Delta \delta_m \sim 10^{-7} M_\odot$.
The median of the optimized values of the loss function is ${\rm loss}(\lambda_{i}) \sim {1.5} \times 10^{-10}$. 
This corresponds to an identical reconstruction up to numerical errors.

\begin{figure*}[]
   \centering
      \includegraphics[width=0.81\textwidth]{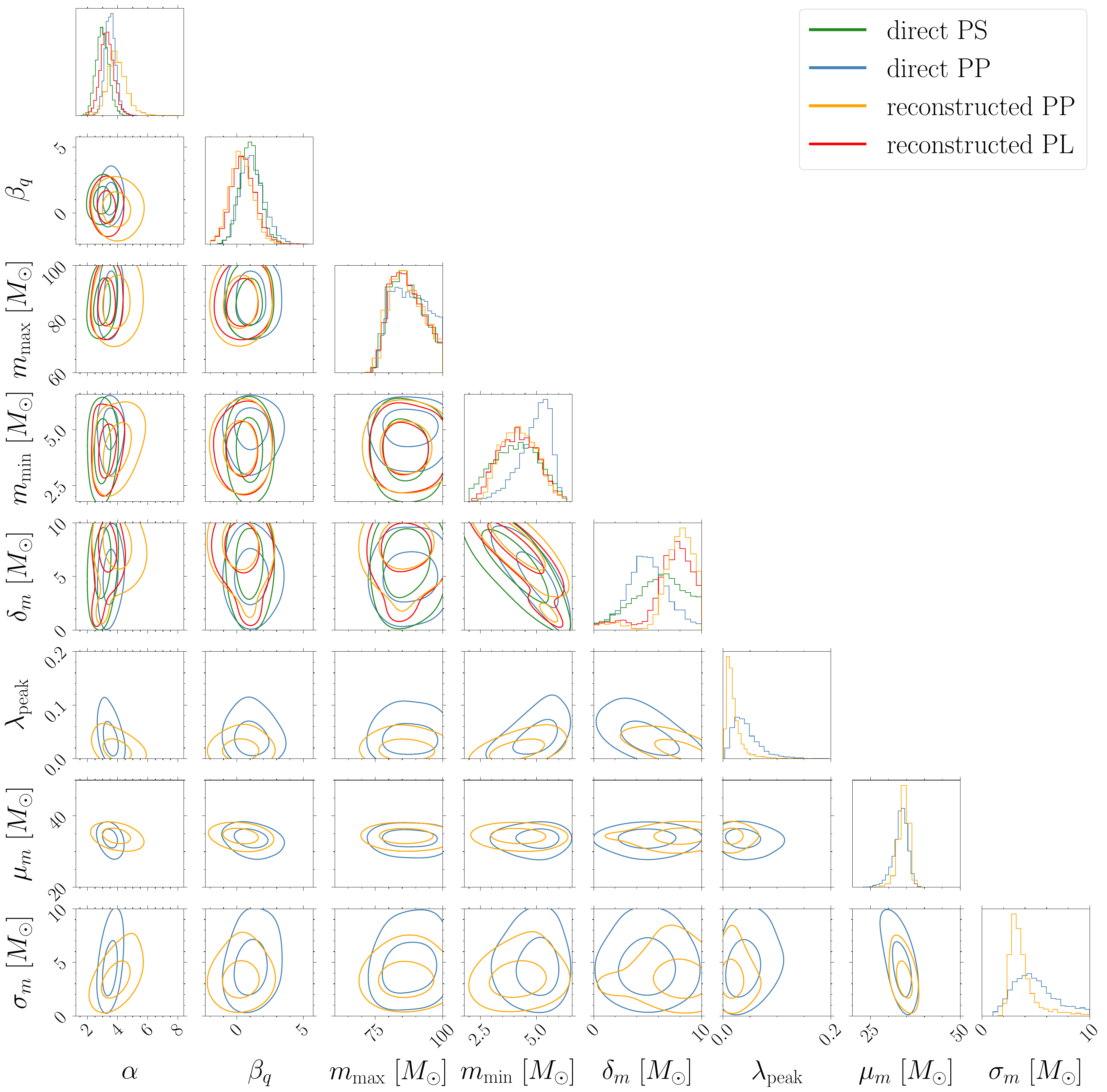} 
  \caption{Reconstruction of the PP (orange) and PL (red) hyperposterior distribution using PS samples obtained from fitting GWTC-3 data. The input PS distribution is shown in green; for comparison, the direct PP fit is shown in blue. Contour levels refer to $90\%$ and $50\%$ credible intervals.
  }
   \label{bigcorner}
\end{figure*}

\begin{figure*}[]
   \includegraphics[width=0.95\columnwidth]{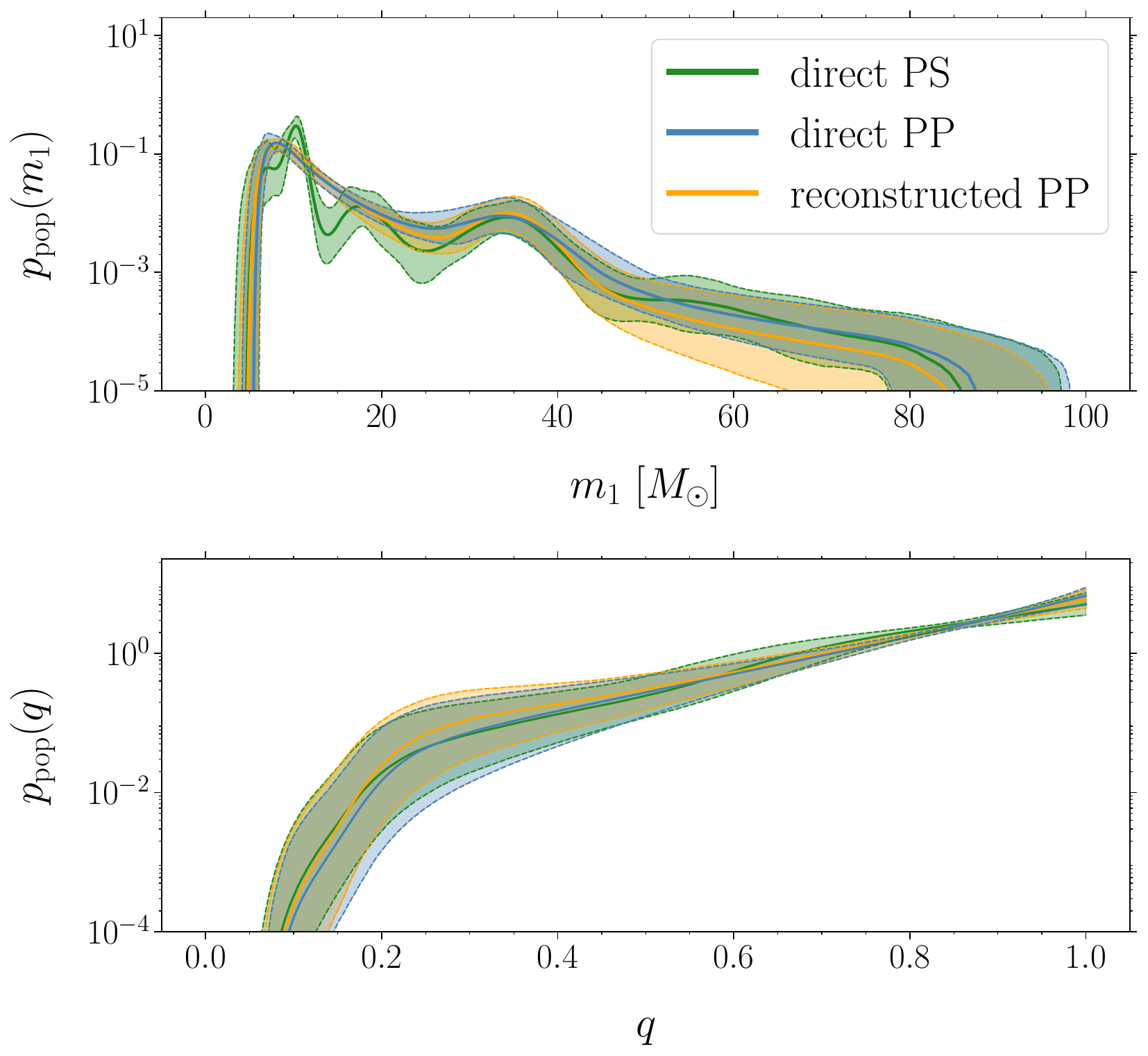}
   $\qquad$
\includegraphics[width=0.95\columnwidth]{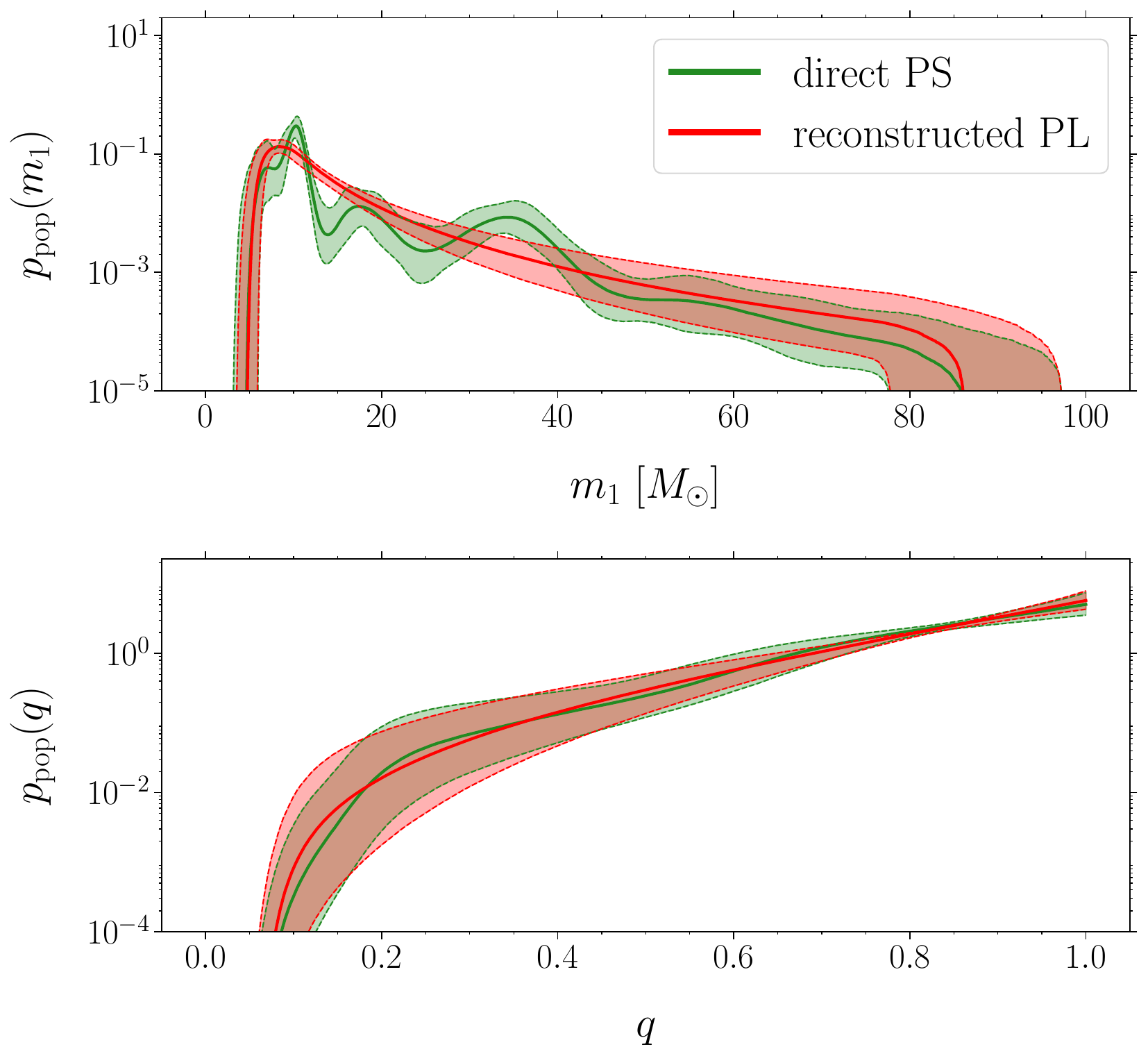} 
  \caption{Reconstruction of the PP (left) and PL (right) population fits using PS samples. The input PS model obtained from fitting GWTC-3 data is shown in green; our reconstructed PP (PL) distribution is shown in orange (red); for comparison, the direct PP fit is shown in blue. %
 The top (bottom) panels show the marginal normalized distributions for the primary mass $m_1$ (mass ratio $q$). Solid curves and shaded regions indicate medians and $90\%$ credible regions, respectively.
 }
   \label{ppopplot}
\end{figure*}

\subsection{\textsc{Power Law+Spline} $\to$ \textsc{Power Law+Peak}}
\label{PS2PP}
Figures~\ref{bigcorner} and \ref{ppopplot} presents our reconstruction of the PP model starting from PS hyperposterior samples. Results are compared against those of Ref.~\cite{2023PhRvX..13a1048A}, where both models were fitted directly to the data. For this optimization, we report a fraction  $\ssim 15\%$ of the input populations that cannot be reproduced with the output model (meaning the optimization resulted in a solution coincident with the imposed boundaries) and have thus been removed to provide a fair comparison; see Sec.~\ref{validity}.
Some hyperparameters are shared, at least nominally, by the input and output models. These are the spectral index of the primary mass $\alpha$, the spectral index of the mass ratio $\beta_{q}$, the smoothing length at low masses $\delta_{m}$, and the minimum and maximum mass cutoffs $m_{\rm min}$, $m_{\rm max}$. For these hyperparameters only, Fig.~\ref{bigcorner} compares the input PS and the output PP hyperposteriors. %

Our reconstruction of the simpler PP model ``cuts through'' the additional overdensities and underdensities of the PS parametrization (see  Fig.~\ref{ppopplot}), which is the expected behavior.  At the same time, the reconstructed population is similar, but not identical, to that obtained by directly fitting the data. %
As discussed in Sec.~\ref{interpretation}, this is to be expected as they are two different, albeit approximately similar, distributions: the direct case $p(\lambda| d)$ and the indirect case $p(\lambda | d, {\rm Rec}_{\gamma})$. %
The reconstructed hyperposterior partially inherits the features of the input model via the fitted hyperparameters $\gamma$. In principle, this might include some of its systematic errors (e.g., overfitting), if present, and samples are weighted by the likelihood of the fitted model rather than the output model. Despite these limitations, the recovered distributions are very consistent with the directly fitted distributions, showing significant overlap in support and similar maximum-a-posteriori parameters. %

In this example, differences are most evident in the hyperposterior distribution of the power-law index $\alpha$ of the primary mass (Fig.~\ref{bigcorner}). Quoting medians and 90\% inner quantiles, the direct PP fit has $\alpha=3.5_{-0.6}^{+0.6}$ while our reconstruction returns $\alpha=3.9_{-0.9}^{+1.1}$ (i.e. steeper decreasing power laws). %
This behavior is driven by overdensities at low masses $5 M_{\odot}\lesssim m_1 \lesssim 15 M_{\odot}$ in the input PS model. 
The PP model does not have the same flexibility, and the adopted priors on the Gaussian moments prevent using the Gaussian component to describe overdensities centered at masses lower than $20 M_{\odot}$. %
This means that the PP model can only capture such low-mass features with a steeper power-law component.
The spectral index of the mass ratio shows a similar behavior, with the reconstructed model returning a lower value $\beta_q=0.3_{-1.3}^{+1.4}$ compared to the direct PP result $\beta_q=1.1_{-1.3}^{+1.7}$.%

\begin{figure*}[]
   \centering
   \includegraphics[width=0.95\columnwidth]{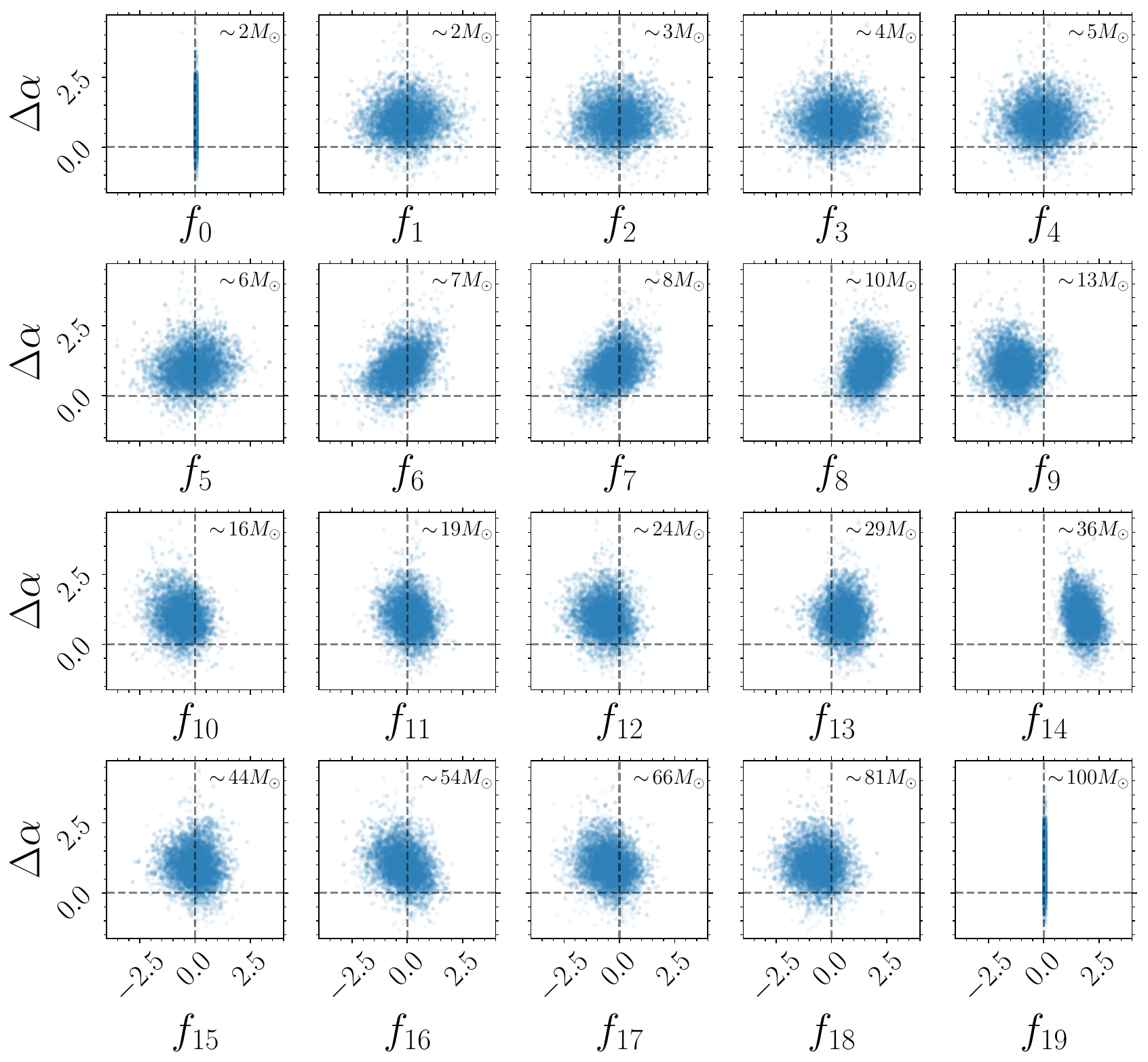} 
   $\qquad$
\includegraphics[width=0.95\columnwidth]{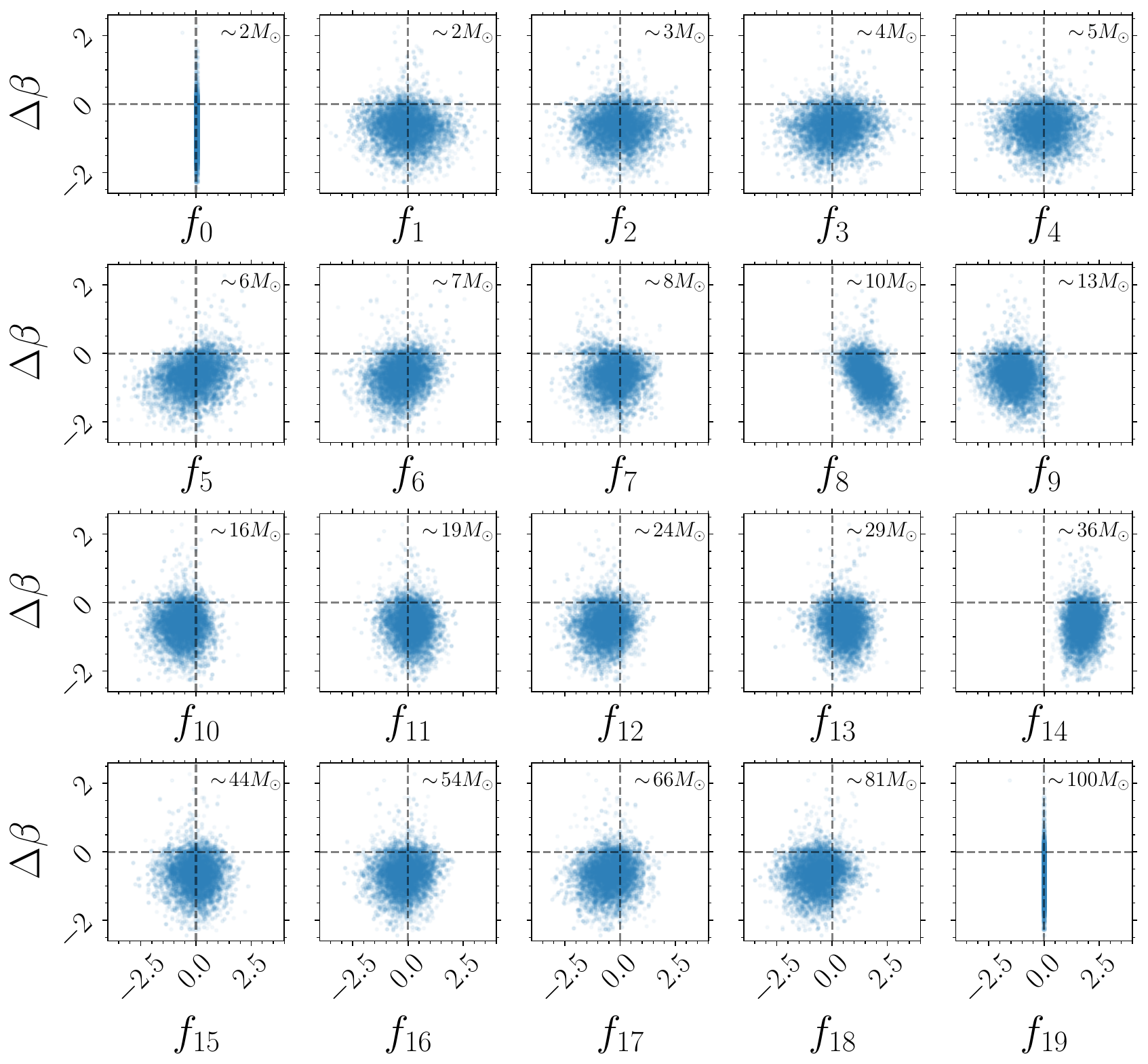} 
  \caption{Correlations in the PS$\to$PP reconstruction. We show joint distributions of the spline perturbation amplitudes and the sample-by-sample difference between output and input spectral indices for the primary mass (left panels) and the mass ratio (right panel). %
Vertical dashed lines indicate locations where the perturbation amplitude is zero;  horizontal dashed lines correspond to identical output and input hyperparameters. By construction, the spline perturbations are set to zero at the first and last nodes. The value of the primary mass at the corresponding spline node is shown in the upper right corner~of~each~subpanel.}
   \label{pert}
\end{figure*}

The intuitive argument we just presented is supported by strong correlations between some of the spline perturbation parameters and the power-law spectral index. We compute  the sample-by-sample deviation $\Delta\alpha$ between the input to output distribution %
and find this is correlated with the amplitude of the spline perturbation amplitude $f_6$ located at $m_1\ssim7M_{\odot}$ %
in the PS model %
(see Fig.~\ref{pert}, left panels).
This indicates that the shift toward steeper power laws in the reconstructed PP model is at least partially connected to that additional degree of freedom of the PS model. Similarly, the spline amplitude $f_{8}$ of the node at 
$m_1\ssim10M_{\odot}$ is anti-correlated with the difference $\Delta\beta_q$ (see Fig.~\ref{pert}, right panels).

Another example of how input features can affect the reconstructed output is shown in the hyperposterior distribution of the minimum mass cutoff (Fig.~\ref{bigcorner}). We reconstruct a PP model with $m_{\rm min}=4.2_{-1.2}^{+1.3} M_\odot$, which is actually in better agreement with the direct PS fit $m_{\rm min}=4.1_{-1.6}^{+1.5} M_\odot$ compared to the direct PP fit $m_{\rm min}=5.1_{-1.5}^{+0.8} M_\odot$. This is precisely due to that extra ${\rm Rec}_{\gamma}$ %
dependency, which contains strong information about such lower $m_{\rm min}$ values being a faithful description of the data. %
The reconstructed hyperposterior of the low-mass smoothing length $\delta_{m}$ is somewhat different from the input (Fig.~\ref{bigcorner}), with the  higher degree of smoothing likely due to the additional low-mass features present in the PS input hyperposterior.

 More broadly, we find that the edges of the distribution are much easier to reconstruct compared to bulk features. The PP-reconstructed posterior distributions of both $m_{\rm min}$ and $m_{\rm max}$ are essentially identical to those of the input PS model (Fig.~\ref{bigcorner}). The minimization algorithm then identifies the optimal solution within the PP parameter space that connects these strongly constrained edges. %

\subsection{\textsc{Power Law+Spline} $\to$ \textsc{Power Law}}

We further explore our procedure by reconstructing the simpler PL model from PS samples. This is a special case where the output model is nested within the input model: the output hyperparameters $\lambda$ are a subset of the output hyperparameters $\gamma$, which then has some additional degrees of freedom. %
In this case, PL corresponds to PS when  all the spline perturbation amplitudes are set to zero $f_{0...19}=0$.

Our reconstruction of the PL model is presented in Figs.~\ref{bigcorner} and \ref{ppopplot}. The fraction of discarded samples here is 8\%, which is smaller than that of the PP case discussed above, corresponding to the intuition that a nested model should be easier to reconstruct.

In general, the PL reconstruction behaves similarly to the PP case described in Sec.~\ref{PS2PP}. 
We find $\alpha=3.3_{-0.7}^{+0.8}$, which is once more higher than that of the PS input $\alpha=3.0_{-0.7}^{+0.7}$, though lower than the PP reconstructed case. 
This is because the power-law distribution in the PL model needs to accommodate not only the overdenisties at lower masses, but also those at $20 - 50M_{\odot}$,  which in PP are instead accounted for with the additional Gaussian component.
Once more, we find that the cutoff hyperparameters are extremely well reproduced; we report $m_{\rm min}=4.2_{-1.3}^{+1.4} M_\odot$ $m_{\rm max}=86.2_{-8.2}^{+11.2} M_\odot$ for the PL reconstruction to be compared with $m_{\rm min}=4.1_{-1.6}^{+1.5} M_\odot$ $m_{\rm max}=86.3_{-8.3}^{+11.0} M_\odot$ for the PS input.

\begin{figure}[]
     \includegraphics[width=\columnwidth]{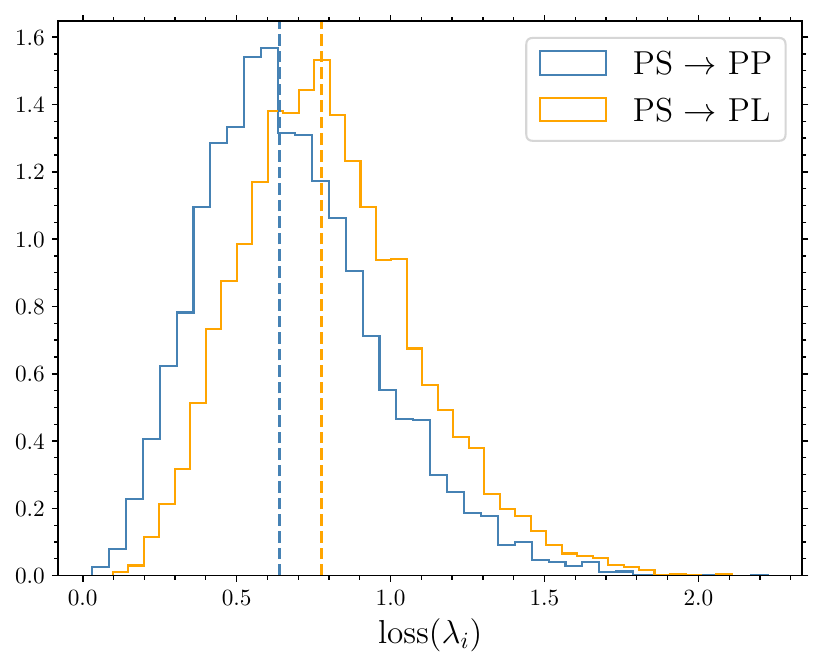} 
     \caption{Values of the loss function at the global minima $\lambda_i$. %
     We compare the reconstruction of the PP (blue) and PL (orange) model using PS as input. Vertical dashed lines indicate the medians of the respective distributions. Lower values of the loss function indicates that the output model better matches the input model, which was in turn fitted to the data.%
     }
   \label{lossfig}
\end{figure}

\subsection{Goodness of fit}
\label{goodness}

The procedure developed in Sec.~\ref{methods} straightforwardly leads to a goodness-of-fit measure, allowing us to assess whether a given reconstruction is appropriate. This is a crucial task when striking the balance between flexibility and interpretability. Simply, a better reconstruction should return lower values of the loss function. %

Figure~\ref{lossfig} shows the values of the minimized loss functions for both the PS$\to$PP and the PS$\to$PL reconstructions. %
We find $\textrm{loss}(\lambda)=0.6_{-0.3}^{+1.2}$ and  $\textrm{loss}(\lambda)=0.8_{-0.4}^{+1.3}$, respectively. The relative distance between the loss for the  PS$\to$PP and PS$\to$PL is $\Delta\textrm{loss}(\lambda)=0.04_{-0.02}^{+0.11}$. That is, the PP model reconstructs the PS model better than PL. Indeed, GWTC-3 contains decisive evidence for additional features in the mass spectrum beyond a simple power law~\cite{2023PhRvX..13a1048A}. This is a very encouraging check, confirming the validity of our method. %
Understanding how to interpret these loss distributions as a quantitative statement about the quality of the model as a description of the population requires an extensive simulation study and should be the subject of future research.

\section{Conclusions}

When fitting experimental data, it is often necessary to test multiple models based on different physical ingredients. %
GW astronomy is no exception: in the context of population analyses, %
different choices provide varying levels of insight, from expressive models that closely capture the underlying astrophysics but may produce model-dependent results, to nonparametric approaches that offer flexibility but involve abstract parameters.
Progress in the field relies on testing multiple population models $p_{\rm pop}(\theta | \lambda)$ to determine which features are robust~\cite{2023PhRvX..13a1048A}. 
This paper offers a practical framework to facilitate this procedure.

While $p_{\rm pop}(\theta | \lambda)$ changes, the other key ingredient of a population fit, namely the detectability function
$p_{\rm det}(\theta)$~
\cite{2019MNRAS.486.1086M,2022hgwa.bookE..45V}, does not, because it only depends on the detector and the data realization. %
The strategy presented here involves first converting the observed set of events into the intrinsic population of sources, with the goal of \emph{describing} the data as accurately as possible. These results can then be postprocessed for \emph{interpreting} the data given preferred astrophysical assumptions. The reconstruction procedure prototyped in this paper relies on a sample-by-sample optimization using the Jeffrey divergence as a loss function. While the notion of distance in probability space is not unique, this is a concrete choice that yields stable results.

Compared to re-fitting the data from scratch, the key advantage of our post-processing strategy is that it does {not} require modeling selection effects. This operation is left entirely in the hands of those who fitted the GW data first. This has immediate applications in our field, with the LVK Collaboration releasing public data products that are then used by a wider community \cite{2023ApJS..267...29A}. %

As with all data analysis procedures, the adopted method becomes part of one's assumptions, entering the resulting hyperposterior as an additional condition: $p(\lambda | d, {\rm Rec}_{\gamma})$. In particular, one should bear in mind that this reconstructed distribution is not the same as the direct hyperposterior of the output model $p(\lambda | d)$. It is important to remember that \emph{all} probabilistic statements rely on numerous implicit assumptions~\cite{jaynes2003probability}. For instance, in the GW population context, these include search strategies, sampling methods, detector sensitivity modeling, and more. Reconstructing population fits as proposed here introduces another assumption, denoted as ${\rm Rec}_{\gamma}$, that must be kept under control. 

For this reason, the input model used in the reconstruction process should provide a faithful description of the data
by using a sufficiently flexible approach,
while taking care to avoid overfitting.
The test case presented here employs the popular PP and PS models ---a choice motivated by the availability of public hyperposterior samples. We speculate that some of the differences observed between the direct PP fit and our PS$\to$PP reconstruction might %
be due to the semi-parametric nature of the input model. While flexible, the PS model still enforces a specific functional form for the mass spectrum (specifically, a spline perturbation to a smoothed power law distribution).
However, this is a somewhat unavoidable uncertainty because no model has infinite capacity in practice. %

Once optimized, the values of the loss function serve as a goodness-of-fit measure. Again, the added condition ${\rm Rec}_{\gamma}$ plays a crucial role here: the distribution of loss values does not indicate which output best fits the data but rather which output best reconstructs the input model, which itself was originally fitted to the data. Additional insights could be gained by employing posterior-predictive checks \cite{2022PASA...39...25R,2020ApJ...891L..31F,2024PhRvD.109j4036M}, which we plan to investigate in future work.

While some of our specific choices may be refined (e.g., the adopted notion of distance and the role of failed optimization attempts),
this paper provides a %
useful tool to perform astrophysical population fits, especially as the number of GW detections continues to grow.
Separating \emph{description} from \emph{interpretation} of data is
a promising way forward for GW population analyses.
Our study represents a first attempt in this direction.

\acknowledgments

We thank 
Walter Del Pozzo, 
Stephen Green, 
Jack Heinzel, 
Christopher Moore, 
Costantino Pacilio, 
Stefano Rinaldi, 
Laura Sberna, 
and the participants of the \textit{``Emerging methods in gravitational-wave population inference''} workshop (IFPU, Trieste) for discussions.
We made use of data from the Gravitational Wave Open Science Center by the LIGO-Virgo-KAGRA Collaboration.
C.M.F., D.G., and A.T. are supported by 
ERC Starting Grant No.~945155--GWmining, 
Cariplo Foundation Grant No.~2021-0555, 
MUR PRIN Grant No.~2022-Z9X4XS, 
MUR Grant ``Progetto Dipartimenti di Eccellenza 2023-2027'' (BiCoQ),
and the ICSC National Research Centre funded by NextGenerationEU. 
C.M.F. is supported by an Anne McLaren Fellowship at the University of Nottingham.
D.G. is supported by MSCA Fellowships No.~101064542--StochRewind and No.~101149270--ProtoBH.
M.M. is supported by the LIGO Laboratory through NSF Grants No.~PHY-1764464 and No.~PHY-2309200.
Computational work was performed at CINECA with allocations 
through INFN, Bicocca, and ISCRA Project No. HP10CW4QH3.

\bibliography{reconstruct}

\end{document}